# Leveraging Deep Learning for Abstractive Code Summarization of Unofficial Documentation

AmirHossein Naghshzan[a,*], Latifa Guerrouj[a], Olga Baysal[b]

[a]*Ecole de Technologie Superieure, 1100 Notre-Dame St W Montreal, H3C1K3, QC, Canada*
[b]*Carleton University, 1125 Colonel By Dr Ottawa, K1S5B6, ON, Canada*

---

## Abstract

Usually, programming languages have official documentation to guide developers with APIs, methods, and classes. However, researchers identified insufficient or inadequate documentation examples and flaws with the API's complex structure as barriers to learning an API. As a result, developers may consult other sources (*e.g.*, StackOverflow, GitHub, etc.) to learn more about an API. Recent research studies have shown that unofficial documentation is a valuable source of information for generating code summaries. We, therefore, have been motivated to leverage such a type of documentation along with deep learning techniques towards generating high-quality summaries for APIs discussed in informal documentation.

This paper proposes an automatic approach using the BART algorithm, a state-of-the-art transformer model, to generate summaries for APIs discussed in StackOverflow. We built an oracle of human-generated summaries to evaluate our approach against it using ROUGE and BLEU metrics which are the most widely used evaluation metrics in text summarization. Furthermore, we evaluated our summaries empirically against a previous work in terms of quality. Our findings demonstrate that using deep learning algorithms can improve summaries' quality and outperform the previous work by an average of %57 for Precision, %66 for Recall, and %61 for F-measure, and it runs 4.4 times faster.

---

*Corresponding author.
*Email addresses:* amirhossein.naghshzan.1@ens.etsmtl.ca (AmirHossein Naghshzan), latifa.guerrouj@etsmtl.ca (Latifa Guerrouj), a.baysal@carleton.ca (Olga Baysal)

## 1. Introduction

During software development, programmers need source documentation to gain information about different APIs, methods, or classes to understand how to use/implement them. Even though developers mostly rely on source code and official documentation as their primary source of information, previous studies have shown that official documentation is not always the best way to extract information about API methods since it may be lengthy and time-consuming and even, in some cases, lacks completeness (Ponzanelli et al., 2015; Uddin and Khomh, 2017). Additionally, Parnin and Treude (2011) has shown that social media plays an essential part in software documentation, and media such as blog postings with tutorials and personal experiences can achieve a high degree of coverage. Overall, developers leverage unofficial documentation including blogs (Parnin and Treude, 2011), StackOverflow (Kavaler et al., 2013), GitHub (Aggarwal et al., 2014) or other sources to get the needed information for their tasks. To help developers, we suggest a novel approach that automatically generates summaries for APIs methods discussed on unofficial documentation, *i.e.*, StackOverflow based on the natural languages around them and their surrounding context.

This approach extends our previous research (Naghshzan et al., 2021). In this research work, we used the TextRank (Mihalcea and Tarau, 2004) algorithm to generate extractive summaries for Android methods discussed in StackOverflow. By conducting a survey with professional developers, we have found that programmers agree on the fact that the generated summaries can be helpful during their development and that context-aware summarizers are needed and can be used as a plugin tool for their daily tasks. This paper focuses on improving the quality of the already generated summaries in (Naghshzan et al., 2021). While we have generated extractive summaries in our previous work, our novel approach generates abstractive summaries. The extractive summarization approach involves choosing key phrases, paragraphs, and sections from the original content and concatenating them into a shorter version (Moratanch and Chitrakala, 2017). While abstractive summarization reproduces key content in a new way after interpreting and examining the text, it uses advanced natural language approaches to construct a new text that delivers the most critical information from the original (Hsu et al., 2018; Moratanch and Chitrakala, 2016). In effect, researchers such as Hsu et al. (2018) have found that abstractive summaries can be more coherent and concise than extractive summaries. We, therefore, investigate the abstractive approach when generating summaries for API methods using our novel deep learning-based approach.

Like in our previous research, we have selected StackOverflow as unofficial documentation and Android APIs methods for summarization. However, we use the BART algorithm to automatically generate summaries for methods to answer the research question mentioned above. This algorithm is one of the state-of-the-art algorithms in summarization, and recent comparisons on CNN and DailyMail datasets have shown that BART outperforms all existing work (Lewis et al., 2020). In addition, we have built an oracle of human-generated summaries to compare with our automatically generated summaries. We have used ROUGE and BLEU metrics to evaluate the accuracy of the generated summaries by calculating Precision, Recall, and F-measure and compared it against recent research as a baseline (Naghshzan et al., 2021).

The main research questions that we address in this journal paper are as follows:

*RQ1: Can we leverage the state-of-the-art deep learning summarization algorithm to automatically generate summaries for API methods discussed on unofficial documentation?*

*RQ2: Does the deep learning algorithm improve the quality of previously generated summaries?*

The purpose is to guide and support developers throughout their tasks by assisting them in rapidly understanding the API methods that are involved in their daily engineering tasks. The findings of this study can be translated into valuable tools such as IDE plugins or recommender systems that both researchers and practitioners can utilize in their practical contexts.

## 2. Technical Background

This section describes the algorithms and techniques applied in our research, BART, ROUGE, and BLEU. We have selected the BART algorithm for the automatic generation of summaries as it outperforms all previous algorithms leveraged for summarization tasks (Lewis et al., 2020). The ROUGE and BLEU metrics are used to evaluate the generated summaries and compare the results against the baseline which, in the recent work related to our study (Naghshzan et al., 2021). To be able to compare our findings with the baseline, we have used the same dataset for comparison, *i.e.*, 3,084,143 Android posts extracted from StackOverflow.

## 2.1. BART

BART stands for Bidirectional Auto-Regressive Transformers is a transformer model developed by Facebook AI research (Lewis et al., 2020) for NLP tasks. The original Transformers, like Google's BERT (Devlin et al., 2019) is a sequence-to-sequence model based on an encoder-decoder architecture. Input and output of the model are sequences, *e.g.*, text with the encoder learning a high-dimensional representation of the input, which is subsequently mapped to the output by the decoder (Vaswani et al., 2017). This is appropriate for classification problems where you may use information from the entire sequence to make a prediction. However, it is less suitable for text production problems in which the prediction is based only on the preceding words. On the other hand, text generation models, such as OpenAI's GPT (Brown et al., 2020), are pre-trained to predict the following token given the preceding sequence of tokens. This pre-training aim produces models suitable for text generation but not classification. As shown in Figure 1, BART is a transformer encoder-encoder (seq2seq) model that combines a bidirectional encoder with an autoregressive. BART is a language model that can be used for language generation, and it has been applied for multi-document summarization tasks (Ma et al., 2022). BART pre-trains itself by corrupting text using a random noise function and then learning a model to recover the original text.

In the original paper (Lewis et al., 2020), the authors reported results on two summarizing datasets, CNN/DailyMail and XSum, which have different characteristics, to compare with the state-of-the-art in summarization. BART outperforms all existing works on the CNN/DailyMail database, which consists of extractive summaries. Also, on the XSum dataset, which is highly abstractive, BART exceeds the best recent work that uses BERT by around 6.0 points across all ROUGE criteria. Benchmark datasets evaluated the model, *i.e.*, CNN, Daily Mail, and an extensive collection of text-summary combinations (Lewis et al., 2020).

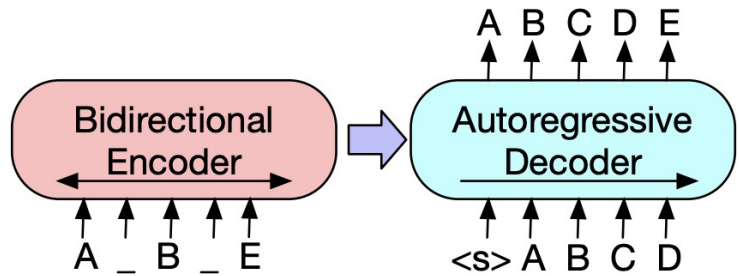

Figure 1: Schematic overview of BART. Extracted from Lewis et al. (2020). A bidirectional model encodes the corrupted document (left), and an autoregressive decoder calculates the likelihood of the original content (right).

### 2.2. TextRank

TextRank is a graph-based unsupervised extractive summarization approach that Mihalcea and Tarau proposed in their widely-cited study (Mihalcea and Tarau, 2004). TextRank technique was inspired by Google's well-known PageRank algorithm (Page et al., 1998) to identify representative key terms in a given dataset (Uddin and Khomh, 2017). TextRank has been shown to create high-quality text summaries, according to Page et al. (1998). This approach has two significant steps: text similarity and the PageRank algorithm. It scores the sentences by calculating co-similarity and producing a weighted graph as a result. The sentences with high weight will be retrieved as a summary (Huang et al., 2020). Figure 2 shows an overview of this algorithm.

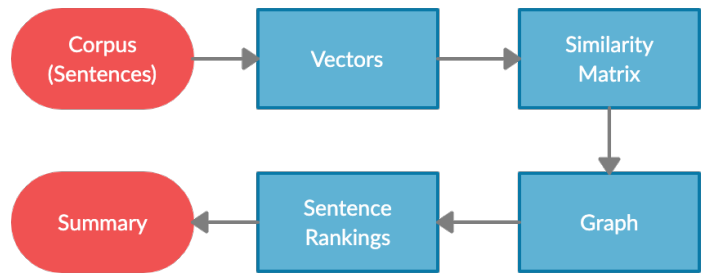

Figure 2: Schematic overview of TextRank. Extracted from Naghshzan et al. (2021).

The algorithm also can be used by selecting words instead of sentences.

A word's importance is determined by the number of votes it receives and the importance of the other words that vote for it, which is similar to how the PageRank algorithm works. If a word is connected to another word through an edge, it casts a vote for the latter, and the importance of the vote is determined by the importance of the first word (Li and Zhao, 2016). As mentioned in the original paper (Mihalcea and Tarau, 2004) TextRank is effective because it considers information recursively gleaned from the entire text (graph) rather than just the local context of a text unit (vertex).

### 2.3. ROUGE

ROUGE or Recall-Oriented Understudy for Gisting Evaluation (Lin, 2004) is the most widely used subject evaluation metric in text summarization (Barbella and Tortora). ROUGE is a set of metrics for evaluating the automatic summarization of texts and machine translations. It compares an automatically generated summary or translation to a list of reference summaries. It focuses on the overlapping of n-grams between the system and human summaries, regardless of semantic or syntactic accuracy (Barbella and Tortora).

### 2.4. BLEU

BLEU, or the Bilingual Evaluation Understudy, is a metric for comparing a candidate's translation of the text to one or more reference translations (Papineni et al., 2002). Although it was designed for translation, it may also be used to analyze text output for natural language processing tasks like paraphrasing and text summarising. BLEU score was highly correlated with human judgments on the translated texts from natural-language tools (Papineni et al., 2002). Despite BLEU's different shortcomings (Qin and Specia, 2015), it remains a popular automated and inexpensive metric to evaluate the quality of translation models (Tran et al., 2019). Assuming that human translations accurately capture the original text's meaning, the BLEU score employs a simple way of measuring the n-grams that overlap in the system translation and the supplied human translations (Qin and Specia, 2015).

## 3. Related Work

Automatic source code summarization has received attention recently since it is an interesting topic for researchers. However, they were mainly concerned with producing a summary for a code block, not an API, and a few researchers have leveraged unofficial documentation in the context of summarization task (Naghshzan et al., 2021; Guerrouj et al., 2015). Moreover, recently deep learning has been widely used for summarization and text generation tasks especially using Transformer models (Vaswani et al., 2017). In this research, we focus on both topics and the most relevant research to our study.

Alambo et al. (2020) proposed an unsupervised multi-document summarization framework. This framework consists of both extractive and abstractive summarization. They used clustering followed by an enhanced multi-sentence compression algorithm to generate informative and relevant summaries for extractive summarization. They used the GPT and the BART algorithms for abstractive summarization. The ROUGE metric was used to evaluate the extractive summaries, and for the abstractive summaries, the authors used five human evaluation metrics, *i.e.*, entailment, coherence, conciseness, readability, and grammar (Alambo et al., 2020). They evaluated their approach using the DUC-2004 benchmark dataset and scientific articles from MAG. Their research findings have shown that the approach matches state-of-the-art extractive summarization and performs better for abstractive summarization. Like this study, we have used the BART algorithm as our primary approach for generating abstractive summaries however, we have selected unofficial documentation *i.e.*, StackOverflow as our source.

Pang et al. (2021) tried to provide abstractive summaries for a given cluster of articles. The researchers used the PEGASUS model and automatic metrics (*i.e.*, ROUGE) and human evaluation to evaluate their approach. The evaluation showed better article summary and cluster-summary entailment in generated summaries compared to other baselines. Unlike this study, we used the BART algorithm for summarization tasks and leveraged StackOverflow instead of academic papers to generate our corpus.

Su et al. (2020) presented CAiRE-COVID, a real-time QA and multi-document summarization system. The system proposes query-focused abstractive and extractive summarization methods to provide relevant information related to the query, a typed question in this case. For the extractive summarization, the authors generated sentence-level representation and used the cosine similarity function to score the sentences. The abstractive summarization part is based on the BART algorithm and the system evaluated on CovidQA dataset (Tang et al., 2020). The authors used ROUGE as the evaluation metric for the performance comparison. The results showed that the ROUGE score improved for extractive and abstractive summarization com-pared to other baselines. In our research, we also used BART for abstractive summarization and ROUGE as one of the metrics of evaluation however, the context in our case is code summarization, and our dataset consists of StackOverflow's Android posts.

To extract a cohesive and informative summary from a single document, (Wu and Hu, 2018) suggested a Reinforced Neural Extractive Summarization model (RNES). Empirical results have indicated that the proposed model can balance the coherence and relevance of sentences and attain state-of-the-art performance on well-known datasets (*i.e.*, CNN and DailyMail). Although we share the same goal of summarization, in our study, we have used an abstractive approach instead of extractive summarization, and the results have been evaluated by ROUGE and BLEU metrics.

Song et al. (2019) proposed an abstract text summarization framework based on LSTM-CNN algorithms that can generate new sentences by investigating finer-grained pieces than sentences, specifically semantic phrases. They applied the framework to CNN and DailyMail datasets. In terms of semantics and syntactic structure, the ATSDL framework outperforms state-of-the-art models in both semantics and syntactic structure. The approach used in this research was state-of-the-art for that time however, the transformers model could outperform them in recent years, and in our study, we have used BART, which is a transformer model.

Kim et al. (2019) tried to predict the success of a movie based on a textual summary of the movie using three deep learning models: an embedding model (ELMo), a one-dimensional convolutional neural network model (1D CNN), and a Long short-term memory (LSTM) network. The evaluation findings demonstrate that the deep learning models perform better than the baseline for the majority class and the ELMo model outperforms CNN and LSTM models. The authors tried three different algorithms to compare the results, while in our case, we have used BART, which outperformed other algorithms in summarization tasks (Lewis et al., 2020) to improve the results of a recent work (Naghshzan et al., 2021; Naghshzan, 2022).

Sridhara et al. (2010) have used Natural Language Processing to extract information to summarize Java methods. In their approach, they preprocessed the source code using a Software Word Usage Model (SWUM), which considers structural and linguistic clues. Then, the natural language summary is generated in a format-based model. Thirteen participants manually evaluated the approach's accuracy, content adequacy, and conciseness. Their work demonstrated that the comments generated for Java methods from source code are accurate and do not miss important information in creating comments. Although this research shares the same idea of code summarization with our study, we used the BART algorithm for abstractive summarization and leveraged unofficial documentation to generate our summaries. Moreover, we have used quantitative evaluation *i.e.*, ROUGE and BLEU metrics and statistical tests for evaluating our results instead of conducting a survey.

McBurney et al. (2016) relied on four different approaches to create a list of features from Java source code documentation. The features at the system-level granularity of software tools are described using Natural Language Sentences (NLS) to provide a clear feature. Their summarization approach is divided into feature list tools and textual analysis tools. In feature list tools, the Latent Dirichlet Allocation (LDA) approach generates a list of topics from the preparation sentences list. TextRank and TLDR tools generated a feature list for a Java source code in their textual analysis tools. They conducted two surveys to evaluate the quality and readability of the generated feature list. These evaluations showed that although LDA can find topics that assist programmers in better understanding projects, it is not always viable to characterize these subjects as natural language sentences retrieved from JavaDocs. However, instead of using LDA in our research, we have used a more recent approach *i.e.*, transformers for summarization tasks. Furthermore, we generated abstractive summarization with high coherence and cohesion compared to extractive summaries (Hsu et al., 2018).

To better understand the use of a Java method, Hu et al. (2018a) proposed a new approach to generate code comments from a large code corpus called DeepCom. In their approach, by applying Natural Language Processing (NLP) techniques, some limitations, such as extracting accurate keywords from the methods that are not named correctly, are solved. Their approach generates a multi-sentence comment by analyzing a block of code. However, in our study, we generate a multi-line summary for an API by analyzing the conversations and discussions related to that API.

LeClair et al. (2020) showed that using a graph-based neural network improves BLEU scores by 4.6% over other graph-based approaches and by 5.7% over flattened AST approaches. They evaluated their technique using a dataset of 2.1 million Java method-comment pairs. The result indicated an improvement over four baseline techniques. Similar to this study, we have tried to improve the quality of a recent code summarization study (Naghshzan et al., 2021). In this regard, we have used transformer models since they could prove their effectiveness compared to neural networks in summarization tasks (Lewis et al., 2020).

## 4. Methodology

This section describes the methodology used to address our research question. Our methodology consists of six significant steps, as illustrated in Figure 3.

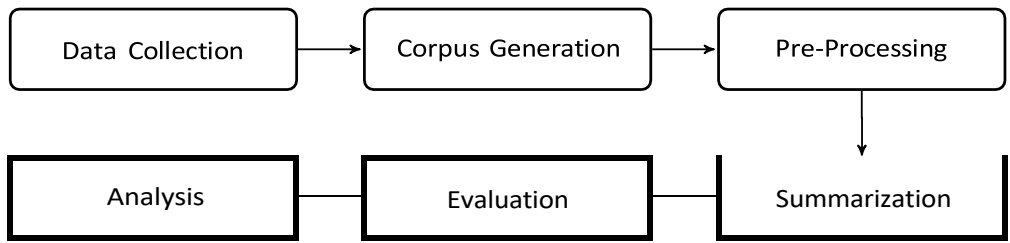

Figure 3: Main phases of the methodology.

As the first step, we have used our previously generated dataset of Stack-Overflow for *Android* posts (Naghshzan et al., 2021). In this section, we briefly explained how the data is gathered. The second step consists of generating a corpus for each API method we aim to summarize. The next step is the cleaning and pre-processing tasks on the corpus. We have applied the pre-trained BART algorithm for the summarization task to automatically generate abstractive summaries. Additionally, We have used the ROUGE and BLEU metrics in the next step to evaluate our summaries and report the obtained findings. In our research's final steps, we analyzed the findings and reported the results.

### 4.1. Data collection

We used the same dataset generated in our previous research (Naghshzan et al., 2021). Furthermore, We leveraged the StackOverflow platform as a type of unofficial documentation and considered only posts with the *Android* tag. We chose Android since it is one of the top eight most discussed topics on StackOverflow based on the type of tags assigned to questions. We have used Stack Exchange API to extract all StackOverflow's questions tagged as *Android*. As reported in Table 1, we have collected 1,266,269 unique Android questions. Then, we extracted all the answers to the collected questions, which resulted in 1,817,874 answers. We have gathered 3,084,143 unique Android posts from StackOverflow, which formed our dataset. We used this dataset to extract API methods from StackOverflow using *Code Snippet* feature introduced by StackOverflow on August 25, 2014.

Table 1: Generated dataset adopted from Naghshzan et al. (2021)

| Posts | Count |
|---|---|
| All Questions | 21,165,633 |
| Android Questions | 1,266,269 |
| Answers to Android Questions | 1,817,874 |
| Total of Android Posts | 3,084,143 |

We utilized the dataset produced in our previous study (Naghshzan et al., 2021). We used StackOverflow as an unofficial documentation source and only considered posts with the *Android* tag. Android was selected because it is one of the top eight most-discussed subjects on StackOverflow based on question tags1. Using the Stack Exchange API, we extracted all Stack-Overflow questions tagged with *Android* from January 2009 to April 2020. According to Table 1, we have gathered 1,266,269 distinct Android queries. Then, we retrieved all the responses to the gathered questions, yielding 1,817,874 responses. Our dataset consists of 3,084,143 unique Android posts collected from StackOverflow. We extracted API methods from StackOverflow using this dataset and the Code Snippet functionality introduced by StackOverflow on August 25, 2014.

In our previous research (Naghshzan et al., 2021), we investigated only fifteen methods for summary generation. The reason was that after the 15th method, we faced a lack of sufficient posts in StackOverflow, and we could not provide enough data for the algorithm to generate high-quality summaries. We previously used TextRank (Mihalcea and Tarau, 2004), an algorithm for generating extractive summaries. However, since researchers have shown that abstractive summaries are more coherent and concise than extractive summaries (Hsu et al., 2018), we leveraged BART, a deep-learning algorithm for generating abstractive summaries.

### 4.2. Corpus Generation

Because each method has its context and its accompanying Stack Over-flow posts, we have generated a corpus for each method. We have chosen answers with a higher score than the average of all Android responses in our sample, which is 2.87. Thus, we have selected StackOverflow responses with a score of at least 3 (scores are integer numbers).

Once the StackOverflow posts have been selected, we have considered their bodies for incorporation into our corpus. However, evaluating all the sentences in the body of a StackOverflow post may not be particularly relevant, as StackOverflow posts may contain irrelevant sentences. Hence, we have considered the following criteria: relevant sentences (criterion 1 and 2) in the context of the summarization task as well as proximity (criteria 3 and 4) already applied in previous works (Naghshzan et al., 2021; Guerrouj et al., 2015; Dagenais and Robillard, 2012; Naghshzan and Ratte, 2023):

1. The opening sentence of each post.
2. The primary sentence contains the method.
3. A sentence that precedes the main sentence.
4. The sentence follows the main clause.

We have used the standard Natural Language Toolkit (NLTK) sentences Tokenizer to split our text into sentences. Following the above procedures, a corpus has been constructed for each API function in our dataset.

### 4.3. Pre-processing

After constructing a corpus for each API method, we followed the procedures below to prepare the corpus for the summarizing algorithm.

We eliminate punctuation, numerals, HTML tags, and other special characters in sentences. We removed stop words from sentences using the Natural Language Toolkit (NLTK). Lemmatization has been applied to appropriately compute words' weight by obtaining their roots. And finally, duplicate sentences have been eliminated from the corpus.

### 4.4. Summarization

Recent context encoders such as GPT, BERT, and BART have been utilized in a large number of NLP research. Since these algorithms are pre-trained on a massive unlabeled corpus, they can provide novel data methods with improved token embeddings. Therefore, developing approaches based on them can obtain enhanced performance (Zhang et al., 2019).

As mentioned earlier, we use BART as the main summarization algorithm for this research. BART is a sequence-to-sequence model trained by corrupting up text with a random noise function and then learning a model to put the original text back together (Lewis et al., 2020). A BART model may take a text sequence as input and create a different text sequence as output. This sort of paradigm applies to machine translation (translating text from one language to another), question-answering (generating responses to a given query based on a specified corpus), text summarization (summarizing or paraphrasing a lengthy text document), and sequence classification (categorizing input text sentences or tokens).

Since BART is an unsupervised pre-trained model, it is possible to fine-tune this language model for a particular NLP task. Pre-training is quite successful for summarizing and even improves content selection power in the absence of copy and coverage procedures (Huang et al., 2020). Since the model has already been pre-trained, enormous labeled datasets are not required for fine-tuning. For this reason, we used the pre-trained BART model developed by (Wolf et al., 2020), which is fine-tuned for summarization. Like a recent study (Hartman and Campion, 2022), transfer learning was done on our dataset for three epochs and four beams with a maximum input token length of 1,024 and a maximum output token length of 50.

The summarization approach consists of two steps. First, we used *BartTokenizer* to make tokens from the text sequences. Similar to the ROBERTa tokenizer (Liu et al., 2019), the BART tokenizer uses byte-level Byte-Pair Encoding. This tokenizer has been trained to consider spaces as tokens. Thus a word will be encoded differently depending on whether it appears at the beginning of the phrase. Secondly we used *BartForConditionalGeneration* to generate summaries. The BART Model with a language modeling head can be used for summarization, so we used this model, which inherits from a pre-trained Model by Wolf et al. (2020). A sample of generated summaries is presented in Table 2 for five investigated methods.

Table 2: Examples of our automatically generated summaries.

| Method | Summary |
|--------|---------|
| activity. onCreate() | OnCreate is the first method in the Activity lifecycle whose job is to create an activity. It is called first time the Activity is created, or when there is a configuration change like a screen rotation. This is the beginning state of an activity lifecycle which its view should be set. Most of the activity initialization code goes here. |
| os.asyncTask. onPostExecute() | This method is called on the thread that created the task. It is called after doInBackground has returned and the result parameter is the value that doIn background has returned. This method is the first method of android activity lifecycle. It helps deliver the calculations that has been done in doInbackground to the main thread. |
| activity. onCreateView() | The Fragment.onCreateView() method is responsible for creating and returning a View. This View is displayed in the UI Fragment, so the desired View must be created in this method. You can return null if the fragment does not provide a UI. The FragmentFragment() method creates a Fragment Fragment and returns a View component that is the root of your fragment's layout. |
| os.asyncTask. doInBackground() | AsyncTask provides a simple way to run tasks in background thread and publish progress and result in the main thread. When creating the task you provide the type of the input parameters, progress, and result with its generic type mechanism. In the doInBackground method you do the long running job and publish the progress. |
| activity. onPause() | In activity's lifecycle when the activity goes from running state to pause state (when we can see the activity but it is not interactable) this method will called. This method implemented, but based on our functions and scenario, we can override it. such as deallocating memory or stop something from processing. Note that when you are on onPause() your Activity is still alive. |

## 5. Empirical Evaluation

In this section, we explain the process of evaluating our automatically generated summaries. First, we built an oracle of human-generated summaries to have a baseline for evaluating the automatically generated summaries to answer our first research question *"Can we leverage the state-of-the-art deep learning summarization algorithm to automatically generate summaries for API methods discussed on unofficial documentation?"*. Then to answer our second research question *"Does deep learning improve the quality of the generated summaries for API methods?"*, we followed a statistical test to compare the quality of our abstractive summaries against the previous extractive approach (Naghshzan et al., 2021).

### 5.1. Building the Oracle

For evaluation purposes, we needed to have reference summaries to be able to compare our generated summaries with them. So, we built an oracle of human-generated summaries for selected APIs and used ROUGE and BLEU metrics to analyze and compare them with our generated summaries. The following sections explain how the oracle was built and the steps taken for analysis.

We built an oracle to generate gold standard summaries for selected APIs. For building the oracle, two annotators contributed to this process, *i.e.*, the first annotator is the first author while the second annotator is non-author of this research. As shown in Figure 4, the process of building the oracle consists of three primary steps: API investigation, summary generation, and summary validation.

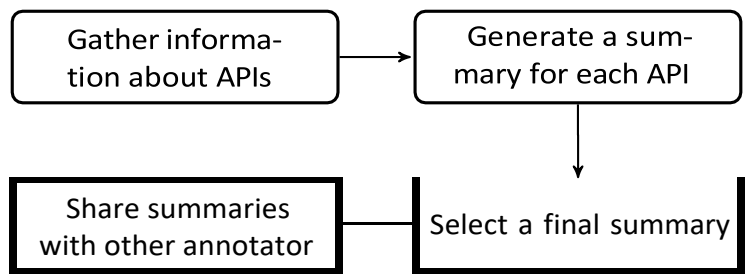

Figure 4: Main steps of building the oracle.

For the first step, each annotator started gathering information about the selected APIs by investigating the official documentation and unofficial sources like StackOverflow, GitHub, Bug Reports, etc., to understand the usage and functionality of each API. When the annotators know enough about an API, the second step is to write a summary for that API based on the gathered data. Each annotator generated a summary for each API based on their understanding of the investigated data. After generating summaries for each API by two annotators individually, the last step is to validate the generated summaries by comparing them and developing the finalized summary. For each API, the annotators shared their generated summary and discussed the main points of each summary. When both annotators agree on the summaries' main points, they generate the final summary by reaching an agreement on the vital points. The final output is a list of summaries for the different studied APIs.

### 5.2. Metrics

By generating the oracle, we have a list of reference summaries to evaluate our generated summaries based on it. In terms of text summarization evaluation, over the years, various evaluation approaches have been used to assess the metrics of summarization (Graham, 2015). However, among all the metrics, ROUGE and BLEU are the most widely-used metrics (Yang et al., 2018; LeClair et al., 2020). BLEU scores are a standard evaluation metric in the source code summarization literature (Hu et al., 2018b; Iyer et al., 2016; LeClair et al., 2019). On the other hand, ROUGE has various unique versions, including eight n-gram counting method options (ROUGE-1; 2; 3; 4; S4; SU4; W; L), binary settings such as word-stemming of summaries, and an option to eliminate or maintain stop-words (Graham, 2015). Hong et al. (2014) proved that ROUGE-2 achieves the strongest correlation with human assessment for summarization system evaluation. This motivated us to use this variant of ROUGE for our evaluation. While BLEU emphasizes n-gram Precision: how much of the generated text appears in the reference text, ROUGE emphasizes Recall: how much of the reference appears in the generated text (LeClair et al., 2020). Therefore like previous studies (Yang et al., 2018; LeClair et al., 2020), we will use a combination of these two metrics to report the evaluation. We will use BLEU for calculating the Precision ($P_l$) and ROUGE for calculating the Recall ($R_l$). Furthermore, F-measure ($F_l$) is the harmonic mean of Precision and Recall (van Rijsbergen, 1979):

$$F_t = \frac{2.P_t.R_t}{P_t+R_t}$$

This provides a credible assessment of the performance of our technique, which depends not only on catching as many words as feasible (Recall) but also on avoiding the output of unnecessary words (Precision). Table 3 reports metrics scores of automatically generated abstractive summaries for 25 Android APIs using the BART algorithm.

Table 3: Metric scores for abstractive summarization using BART.

| Method | Precision | Recall | F-Measure |
|---|---|---|---|
| asyncTask.onPostExecute | 0.9393 | 0.7948 | 0.8611 |
| fragment.onCreateView | 0.6418 | 0.7277 | 0.6820 |
| activity.onCreate | 0.6608 | 0.7448 | 0.7002 |
| asyncTask.doInBackground | 0.6563 | 0.6237 | 0.6396 |
| activity.onPause | 0.4545 | 0.4381 | 0.4461 |
| activity.findViewById | 0.875 | 0.6764 | 0.7630 |
| activity.onDestroy | 0.9324 | 0.6666 | 0.7774 |
| activity.finish | 0.62 | 0.5864 | 0.6027 |
| activity.setContentView | 0.5555 | 0.7963 | 0.6545 |
| activity.startActivityForResult | 0.5 | 0.475 | 0.4871 |
| recyclerview.onBindViewHolder | 0.75 | 0.6153 | 0.6760 |
| activity.startActivity | 0.5530 | 0.6825 | 0.6109 |
| activity.onBackPressed | 0.9318 | 0.7735 | 0.8453 |
| activity.onActivityResult | 0.4042 | 0.6551 | 0.4999 |
| activity.onStop | 0.3902 | 0.3809 | 0.3855 |
| activity.onStart | 0.7741 | 0.8482 | 0.8095 |
| adapter.notifyDataSetChanged | 0.4258 | 0.3905 | 0.4073 |
| adapter.getView | 0.6728 | 0.7231 | 0.6970 |
| view.onDraw | 0.7032 | 0.8451 | 0.7676 |
| activity.onSaveInstanceState | 0.8027 | 0.6459 | 0.7158 |
| activity.onNewIntent | 0.5725 | 0.4286 | 0.4902 |
| activity.onActivityCreated | 0.7394 | 0.8725 | 0.8005 |
| activity.onCreateOptionsMenu | 0.856 | 0.6551 | 0.7421 |
| activity.runOnUiThread | 0.53 | 0.6034 | 0.5643 |
| asyncTask.onProgressUpdate | 0.856 | 0.7348 | 0.7907 |
| **Average** | **0.5718** | **0.6634** | **0.6142** |

### 5.3. Statistical Test

To validate our second research question, we follow the Basili framework to describe our study, which consists of the following parts: definition, planning, operation, and analysis (Basili et al., 1986). An overview of this framework is shown in Figure 5.

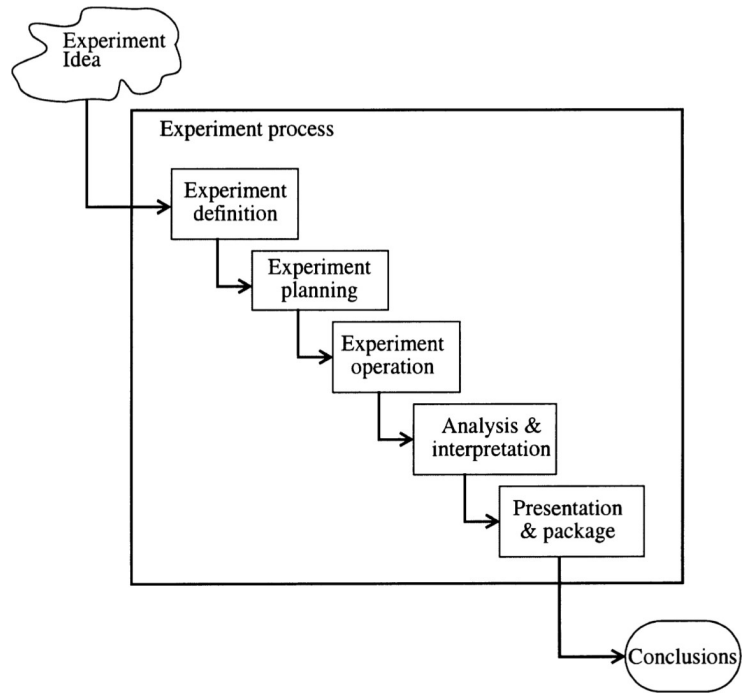

Figure 5: Overview of Basili experiment process. Extracted from Basili et al. (1986)

### 5.3.1. Definition and Planning of the Study

The *goal* of our study is to investigate if we could improve the quality of our automatically generated summaries compared to our previous research (Naghshzan et al., 2021).

The *Quality* focus is represented by the performance of the two investigated algorithms, *i.e.*, TextRank, and BART, measured in terms of, Precision, Recall, and F-measure.

The *Perspective* is of researchers and developers who would like to search for extra information about APIs and methods during their development and software engineering tasks.

The type of algorithms used is the main independent variable in our study. This factor has two treatments, *i.e.*, BART and TextRank.

The dependent variable in our study is the performance of the mentioned algorithms measured in terms of Precision, Recall, and F-measure.

### 5.3.2. Operation

By conducting this empirical evaluation, we investigate three questions derived from our main research question. The questions and the null and alternative hypotheses corresponding to each research question are formulated as follows:

1. ***RQ1**: How does the BART algorithm perform compared to TextRank in terms of Precision?*

   - $H_{0\_1}$: There is no statistically significant difference between BART and TextRank in terms of Precision.
   - $H_{1\_1}$: There is statistically a significant difference between BART and TextRank in terms of Precision.

2. ***RQ2**: How does the BART algorithm perform compared to TextRank in terms of Recall?*

   - $H_{0\_2}$: There is no statistically significant difference between BART and TextRank in terms of Recall.
   - $H_{1\_2}$: There is statistically a significant difference between BART and TextRank in terms of Recall.

3. ***RQ3**: How does the BART algorithm perform compared to TextRank in terms of F-Measure?*

   - $H_{0\_3}$: There is no statistically significant difference between BART and TextRank in terms of F-Measure.
   - $H_{1\_3}$: There is a statistically significant difference between BART and TextRank in terms of F-Measure.

### 5.3.3. Analysis

To compare our BART results against the previous TextRank approach, we calculated the Precision, Recall, and F-measure of summaries generated by the TextRank algorithm. Table 4 reports the calculated scores for 15 Android APIs using the TextRank algorithm.

Table 4: Metric scores for extractive summarization using TextRank.

| Method | Precision | Recall | F-Measure |
|---|---|---|---|
| asyncTask.onPostExecute | 0.3818 | 0.5908 | 0.4638 |
| fragment.onCreateView | 0.1940 | 0.3939 | 0.2599 |
| activity.onCreate | 0.1690 | 0.4 | 0.2376 |
| asyncTask.doInBackground | 0.2413 | 0.3159 | 0.2736 |
| activity.onPause | 0.1666 | 0.2391 | 0.1964 |
| activity.findViewById | 0.1791 | 0.3636 | 0.2399 |
| activity.onDestroy | 0.1408 | 0.3333 | 0.1980 |
| activity.finish | 0.2222 | 0.3636 | 0.2758 |
| activity.setContentView | 0.3030 | 0.5454 | 0.3896 |
| activity.startActivityForResult | 0.2222 | 0.5128 | 0.3100 |
| recyclerview.onBindViewHolder | 0.1941 | 0.4878 | 0.2777 |
| activity.startActivity | 0.2941 | 0.4878 | 0.3669 |
| activity.onBackPressed | 0.4745 | 0.5284 | 0.5578 |
| activity.onActivityResult | 0.274 | 0.3285 | 0.2987 |
| activity.onStop | 0.1281 | 0.1934 | 0.1541 |
| **Average** | **0.2504** | **0.4056** | **0.3096** |

Furthermore, as we extended the number of generated summaries from 15 to 25 APIs in this research, we have used only the same 15 APIs reported in (Naghshzan et al., 2021).

Our method of analysis relies on both descriptive statistics and statistical analyses. We used box plots to present the results for each descriptive statistic (McGill et al., 1978).

We used the Shapiro-Wilk normality test (Shapiro and Wilk, 1965) to determine whether or not our data follows a normal distribution. When the p-value exceeds 0.05, the data distribution is not significantly different from the normal distribution.

The result of the Shapiro-Wilk test is reported in Table 5. The p-value for all performance measures is more than 0.05. Therefore, both algorithms follow a normal distribution. As a result, we must use parametric statistical tests on our data.

Table 5: Results of the Shapiro-Wilk normality test.

| Algorithm | Precision | Recall | F-Measure |
|-----------|-----------|--------|-----------|
| BART      | 0.9536    | 0.9367 | 0.9418    |
| TextRank  | 0.8951    | 0.9615 | 0.9139    |

We compared the Precision, Recall, and F-measure using the paired T-test (Student, 1908) is a parametric test for pair-wise median comparison. The T-test determines whether or not the median difference between the two algorithms is zero. The results are interpreted as statistically significant at $\alpha$ = 5%. Then parametric t-stat and p-value were calculated to determine the magnitude of the effect between two different summarization algorithms. A t-score of 0 indicates that the sample results equal the null hypothesis. As the difference between the sample data and the null hypothesis increases, the absolute value of the t-score increases. Larger t-scores mean more difference between groups, and smaller t-scores mean more similarity between groups. The p-value is the variable that allows us to reject the null hypothesis ($H_0$) or, in other words, to establish that the two groups are different. The p-value greater than $\alpha$ shows a failure to reject the statistical test's null hypothesis. Otherwise, we can reject the null hypothesis of the statistical test.

## 6. Results and Discussion

This section presents the results of our comparison between BART and TextRank algorithms. The descriptive statistics are reported by providing the following box plots. Figures 6, 7 and 8 show that BART could outperform TextRank in terms of Precision, Recall, and F-Measure.

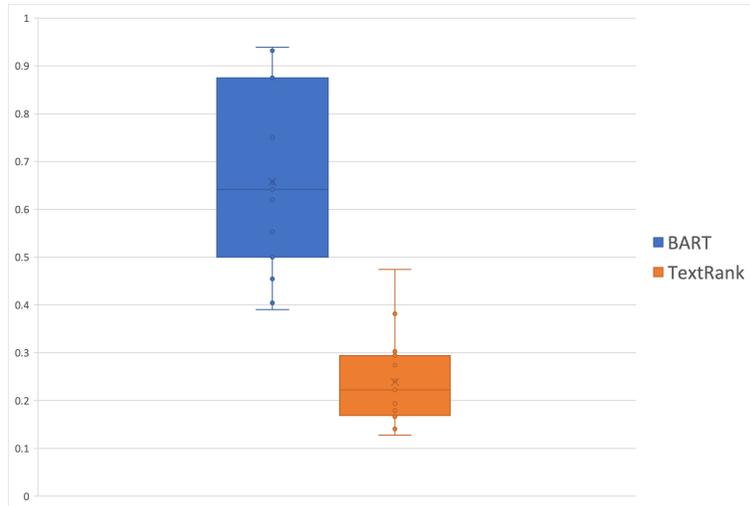

Figure 6: Box-plots of Precision for BART and TextRank algorithms.

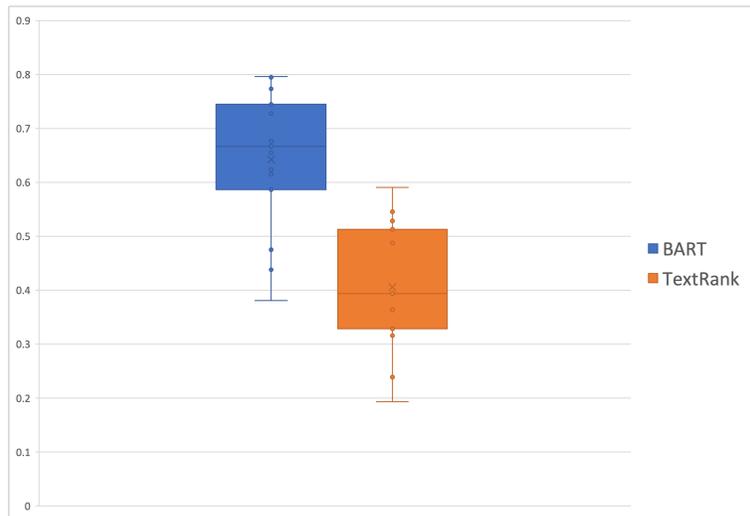

Figure 7: Box-plots of Recall for BART and TextRank algorithms.

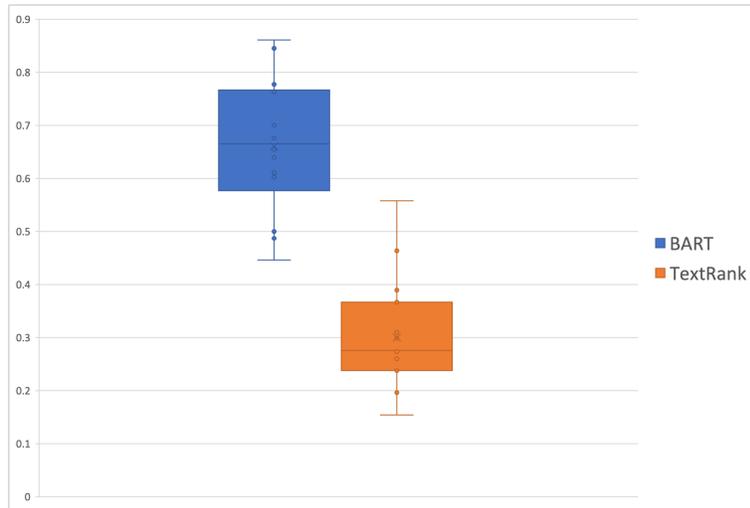

Figure 8: Box-plots of F-Measure for BART and TextRank algorithms.

Table 6: t-scores and -values for comparison between BART and TextRank algorithms.

| Precision | | Recall | | F-measure | |
|---|---|---|---|---|---|
| t-score | p-value | t-score | p-value | t-score | p-value |
| 8.9454 | 3.64E-07 | 9.0516 | 3.16E-07 | 10.9914 | 2.86E-08 |

Furthermore, the statistical result of the T-test is provided in Table 6. As seen in the table, for Precision, the t-score is 8.9454, which is greater than the t-critical, which is 1.761 in this statistical test. This difference shows that the difference, in this case, is significant. Moreover, the p-value is 3.64E-07 which is less than $\alpha = 0.05$, and proves that we reject the null hypothesis of RQ1 in section 5.3.2 and conclude that BART and TextRank have significant differences in terms of Precision.

Regarding Recall, Table 6 shows that the t-score is 9.0516 which is greater than our t-critical (1.761), and also the p-value (3.16#-07) is less than $\alpha = 0.05$. Like the previous statement, we reject the null hypothesis of RQ2 in section 5.3.2 and conclude that BART and TextRank have significant differences in terms of Recall.

Considering the F-Measure, the same argument applies. As shown in Table 6 t-score is 10.9914 which is greater than the t-critical (1.761) and the p-value is 2.86E-08 and it is less than $\alpha = 0.05$. Therefore, we reject the null hypothesis of RQ3 in section 5.3.2 and conclude that BART and TextRank have significant differences in F-measure.

Table 7 shows the execution time of BART and TextRank for each API. The average execution time for BART is 145.64 seconds, while TextRank's average execution time is 641.29 seconds. The results show that BART could perform 4.4x faster compared to TextRank. It is worth mentioning that this time is only for executing algorithms on the data. Since we used a pre-trained model of BART, the training time is not included. This is why BART, a deep learning algorithm that needs training time, runs faster than TextRank, which is an unsupervised learning algorithm.

Table 7: Execution time of BART and TextRank.

| Method | BART(sec) | TextRank(sec) |
|---|---|---|
| asyncTask.onPostExecute | 184.81 | 739.63 |
| fragment.onCreateView | 126.88 | 583.75 |
| activity.onCreate | 269.92 | 1435.85 |
| asyncTask.doInBackground | 113.81 | 437.92 |
| activity.onPause | 123.91 | 585.05 |
| activity.findViewById | 114.37 | 446.65 |
| activity.onDestroy | 123.42 | 547.29 |
| activity.finish | 125.90 | 586.11 |
| activity.setContentView | 123.79 | 485.70 |
| activity.startActivityForResult | 131.08 | 490.42 |
| recyclerview.onBindViewHolder | 136.85 | 459.36 |
| activity.startActivity | 153.03 | 556.74 |
| activity.onBackPressed | 142.70 | 655.53 |
| activity.onActivityResult | 188.41 | 716.73 |
| activity.onStop | 125.33 | 892.73 |
| **Average** | **145.64** | **641.29** |

To wrap up, we conclude that there is a significant difference between BART and TextRank algorithms in terms of Precision, Recall, and F-Measure. As seen in box plots, the BART algorithm outperforms the TextRank algorithm.

To answer our main research questions, regarding our first research question: *Can we leverage the state-of-the-art deep learning summarization algorithm to automatically generate summaries for API methods discussed on unofficial documentation?*, we showed that with an average of 0.5718 for Precision, 0.6634 for Recall and 0.6142 for F-Measure, BART could successfully generate abstractive summaries. These results show that BART is an appropriate algorithm for generating abstractive summaries, and unofficial documentation *i.e.*, StackOverflow has the potential of providing information for APIs and methods discussed in it, and the community can benefit it from extracting information for code summarization tasks.

Considering the second research question: *Does deep learning algorithm improve the quality of previously generated summaries?* We considered a previous study as a baseline and compared our abstractive summaries against it. By rejecting all null hypotheses mentioned in section 5.3.2, we conclude that we could improve the quality of automatically generated summaries by applying the BART algorithm for generating summaries.

While our previous research (Naghshzan et al., 2021) proved that using unofficial documentation as a complementary source for generating summaries is promising, in this study, we could extend the idea by generating high-quality summaries of APIs discussed in Stackoverflow as an example of unofficial documentation. Moreover, we believe the significant difference between BART and TextRank algorithm is not only because BART is a state-of-the-art summarization algorithm but also because BART generates abstractive summaries. As Hsu et al. (2018) suggests, abstractive summaries are more coherent and closer to human-generated summaries than extractive summarization methods like TextRank. This is probably one of the reasons BART got far better results against TextRank.

During our research, we observed that some APIs (*i.e.*, activity.onStop) have low scores in both approaches compared to others. Based on our investigation, we believe the algorithms cannot generate high-quality summaries for APIs with a low number of data in our corpus. For instance for *asyncTask.onPostExecute* 34,562 posts extracted from StackOverflow, while in case of *activity.onStop* we could extract only 3,389 posts. Therefore lack of sufficient data results in summaries that are not close enough to human-generated summaries, especially in the case of extractive summarization since the Tex-

tRank algorithm uses the same sentences from different posts, which may be unrelated.

A vital point to mention is that although the TextRank algorithm has low metric scores compared to BART, our previous empirical survey proved that developers found the summaries helpful during the development steps. We believe this conflict is because of extractive summarization. In extractive summarization, the final summary has low coherence since the sentences are selected from different texts with different contexts. As a result, it would be far from human-generated summaries. However, the TextRank algorithm could extract essential information from the corpus and summarize it so developers could use it. Undoubtedly, having a coherent summary like our BART-generated summaries would be more welcomed by developers and the community since it has the key points and is easier to understand.

## 7. Threats to Validity

Although our study has been empirically and statistically evaluated, there are threats to its validity. The most common ones are covered in the following sections:

### 7.1. External validity

External validity is the degree to which a study's findings are generalizable (Godwin et al., 2004). We selected StackOverflow as our unofficial source and Android programming languages examined in our research. These selections are not necessarily representative of all software programming languages. We considered these selections only for the sake of simplicity. Otherwise, our pipeline is not limited to these selections, and all the steps are extendable to other programming languages and unofficial documentation.

### 7.2. Reliability validity

Reliability is the degree to which a study produces comparable results under different conditions (Godwin et al., 2004). The frame time of investigated posts on StackOverflow is reported in the paper. Furthermore, we used the Python *rouge* package to calculate the ROUGE score and *nltk* package to calculate the BLEU score (with the parameters reported in the paper). Our online Appendix provides additional data that may be required to replicate our study[1].

### 7.3. Conclusion validity

The extent to which a study's conclusions are derived from sufficient data analysis and all research questions are accurately answered is referred to as conclusion validity (Garc´ıa-Pêrez, 2012). We conducted statistical tests to report the findings. We used ROUGE and BLEU, two well-known and widely-used metrics for reporting the score, and combined them to understand our results better.

## 8. Conclusion and Future Work

Code summarization is an interesting domain for NLP researchers since it can help software engineers increase productivity and efficiency during development. However, few studies focused on generating summaries for APIs by leveraging unofficial documentation as a source for summarization.

In this research, we extended our previous study (Naghshzan et al., 2021) on code summarization. We proposed a novel approach to generate summaries for APIs and methods discussed on unofficial documentation based on their surrounding context. Like our previous study, we selected StackOverflow as an unofficial source of summarization and its Android posts as our dataset. However, we applied BART, a state-of-the-art text generation algorithm, as the primary summarization method. Moreover, we evaluated generated summaries of twenty-five methods using ROUGE and BLEU metrics against an oracle of human-generated summaries. Furthermore, we conducted an empirical evaluation to compare our approach against a previous (Naghshzan et al., 2021) study as a baseline.

Our findings show that unofficial documentation, such as StackOverflow, has the potential to be a source for generated summaries. Moreover, our empirical evaluation proves that abstractive summarization using deep learning algorithms can improve the quality of summaries compared to extractive summarization by average of %57 for Precision, %66 for Recall and %61 for F-Measure and it runs 4.4 times faster.

As our future extension, we are interested in combining other types of unofficial documentation such as GitHub, bug reports, etc. Additionally, we intend to extend our data from the Android API methods to other domains and empirically evaluate our approach by conducting a large-scale survey of professional developers and checking whether the summaries are productive in developing steps or not.

---

[1]https://github.com/DataMining2022/summarization

## Acknowledgements

Guerrouj would like to acknowledge the support of the Natural Sciences and Engineering Research Council of Canada (NSERC), RGPIN-4712-2016. Baysal would like to acknowledge the support of the Natural Sciences and Engineering Research Council of Canada (NSERC), RGPIN-2021-03809.

## References

Aggarwal, K., Hindle, A., Stroulia, E., 2014. Co-evolution of project documentation and popularity within github, in: Proceedings of the 11th working conference on mining software repositories, pp. 360–363.

Alambo, A., Lohstroh, C., Madaus, E., Padhee, S., Foster, B., Banerjee, T., Thirunarayan, K., Raymer, M., 2020. Topic-centric unsupervised multi-document summarization of scientific and news articles, in: 2020 IEEE International Conference on Big Data (Big Data), IEEE Computer Society, Los Alamitos, CA, USA. pp. 591–596. URL: https://doi.ieeecomputersociety.org/10.1109/BigData50022.2020.9378403, doi:10.1109/BigData50022.2020.9378403.

Barbella, M., Tortora, G., . Rouge metric evaluation for text summarization techniques. Available at SSRN 4120317 .

Basili, V.R., Selby, R.W., Hutchens, D.H., 1986. Experimentation in software engineering. IEEE Transactions on Software Engineering SE-12, 733–743. doi:10.1109/TSE.1986.6312975.

Brown, T.B., Mann, B., Ryder, N., Subbiah, M., Kaplan, J., Dhariwal, P., Neelakantan, A., Shyam, P., Sastry, G., Askell, A., Agarwal, S., Herbert-Voss, A., Krueger, G., Henighan, T., Child, R., Ramesh, A., Ziegler, D.M., Wu, J., Winter, C., Hesse, C., Chen, M., Sigler, E., Litwin, M., Gray, S., Chess, B., Clark, J., Berner, C., McCandlish, S., Radford, A., Sutskever, I.,

Amodei, D., 2020. Language models are few-shot learners, in: Proceedings of the 34th International Conference on Neural Information Processing Systems, Curran Associates Inc., Red Hook, NY, USA. p. .

Dagenais, B., Robillard, M.P., 2012. Recovering traceability links between an api and its learning resources, in: 2012 34th International Conference on Software Engineering (ICSE), pp. 47–57. doi:10.1109/ICSE.2012.6227207.

Devlin, J., Chang, M.W., Lee, K., Toutanova, K., 2019. BERT: Pre-training of deep bidirectional transformers for language understanding, in: Proceedings of the 2019 Conference of the North American Chapter of the Association for Computational Linguistics: Human Language Technologies, Volume 1 (Long and Short Papers), Association for Computational Linguistics, Minneapolis, Minnesota. pp. 4171–4186. URL: https://aclanthology.org/N19-1423, doi:10.18653/v1/N19-1423.

García-Pérez, M., 2012. Statistical conclusion validity: Some common threats and simple remedies. Frontiers in Psychology 3. URL: https://www.frontiersin.org/article/10.3389/fpsyg.2012.00325, doi:10.3389/fpsyg.2012.00325.

Godwin, M., Ruhland, L., Casson, I., MacDonald, S., Delva, D., Birtwhistle, R., Lam, M., Seguin, R., 2004. Pragmatic controlled clinical trials in primary care: the struggle between external and internal validity. BMC medical research methodology 3, 28. doi:10.1186/1471-2288-3-28.

Graham, Y., 2015. Re-evaluating automatic summarization with BLEU and 192 shades of ROUGE, in: Proceedings of the 2015 Conference on Empirical Methods in Natural Language Processing, Association for Computational Linguistics, Lisbon, Portugal. pp. 128–137. URL: https://aclanthology.org/D15-1013, doi:10.18653/v1/D15-1013.

Guerrouj, L., Bourque, D., Rigby, P.C., 2015. Leveraging informal documentation to summarize classes and methods in context, in: 2015 IEEE/ACM 37th IEEE International Conference on Software Engineering, pp. 639–642. doi:10.1109/ICSE.2015.212.

Hartman, V., Campion, T.R., 2022. A day-to-day approach for automating the hospital course section of the discharge summary, in: AMIA Annual

Symposium Proceedings, American Medical Informatics Association. p. 216.

Hong, K., Conroy, J., Favre, B., Kulesza, A., Lin, H., Nenkova, A., 2014. A repository of state of the art and competitive baseline summaries for generic news summarization, in: Proceedings of the Ninth International Conference on Language Resources and Evaluation (LREC'14), European Language Resources Association (ELRA), Reykjavik, Iceland. pp. 1608–1616. URL: http://www.lrec-conf.org/proceedings/lrec2014/pdf/1093_Paper.pdf.

Hsu, W.T., Lin, C.K., Lee, M.Y., Min, K., Tang, J., Sun, M., 2018. A unified model for extractive and abstractive summarization using inconsistency loss. arXiv preprint arXiv:1805.06266 .

Hu, X., Li, G., Xia, X., Lo, D., Jin, Z., 2018a. Deep code comment generation, in: , Association for Computing Machinery, New York, NY, USA. p. 200–210. URL: https://doi.org/10.1145/3196321.3196334, doi:10.1145/3196321.3196334.

Hu, X., Li, G., Xia, X., Lo, D., Lu, S., Jin, Z., 2018b. Summarizing source code with transferred api knowledge, in: Proceedings of the Twenty-Seventh International Joint Conference on Artificial Intelligence, IJCAI-18, International Joint Conferences on Artificial Intelligence Organization. pp. 2269–2275. URL: https://doi.org/10.24963/ijcai.2018/314, doi:10.24963/ijcai.2018/314.

Huang, D., Cui, L., Yang, S., Bao, G., Wang, K., Xie, J., Zhang, Y., 2020. What have we achieved on text summarization? arXiv preprint arXiv:2010.04529 .

Iyer, S., Konstas, I., Cheung, A., Zettlemoyer, L., 2016. Summarizing source code using a neural attention model, in: Proceedings of the 54th Annual Meeting of the Association for Computational Linguistics (Volume 1: Long Papers), Association for Computational Linguistics, Berlin, Germany. pp. 2073–2083. URL: https://aclanthology.org/P16-1195, doi:10.18653/v1/P16-1195.

Kavaler, D., Posnett, D., Gibler, C., Chen, H., Devanbu, P., Filkov, V., 2013. Using and asking: Apis used in the android market and asked

about in stackoverflow, in: International Conference on Social Informatics, Springer. pp. 405–418.

Kim, Y.J., Cheong, Y.G., Lee, J.H., 2019. Prediction of a movie's success from plot summaries using deep learning models, in: Proceedings of the Second Workshop on Storytelling, Association for Computational Linguistics, Florence, Italy. pp. 127–135. URL: https://aclanthology.org/W19-3414, doi:10.18653/v1/W19-3414.

LeClair, A., Haque, S., Wu, L., McMillan, C., 2020. Improved code summarization via a graph neural network, in: Proceedings of the 28th International Conference on Program Comprehension, Association for Computing Machinery, New York, NY, USA. p. 184–195. URL: https://doi.org/10.1145/3387904.3389268, doi:10.1145/3387904.3389268.

LeClair, A., Jiang, S., McMillan, C., 2019. A neural model for generating natural language summaries of program subroutines, in: Proceedings of the 41st International Conference on Software Engineering, IEEE Press. p. 795–806. URL: https://doi.org/10.1109/ICSE.2019.00087, doi:10.1109/ICSE.2019.00087.

Lewis, M., Liu, Y., Goyal, N., Ghazvininejad, M., Mohamed, A., Levy, O., Stoyanov, V., Zettlemoyer, L., 2020. BART: Denoising sequence-to-sequence pre-training for natural language generation, translation, and comprehension, in: Proceedings of the 58th Annual Meeting of the Association for Computational Linguistics, Association for Computational Linguistics, Online. pp. 7871–7880. URL: https://aclanthology.org/2020.acl-main.703, doi:10.18653/v1/2020.acl-main.703.

Li, W., Zhao, J., 2016. Textrank algorithm by exploiting wikipedia for short text keywords extraction, in: 2016 3rd International Conference on Information Science and Control Engineering (ICISCE), IEEE. pp. 683–686.

Lin, C.Y., 2004. ROUGE: A package for automatic evaluation of summaries, in: Text Summarization Branches Out, Association for Computational Linguistics, Barcelona, Spain. pp. 74–81. URL: https://aclanthology.org/W04-1013.

Liu, Y., Ott, M., Goyal, N., Du, J., Joshi, M., Chen, D., Levy, O., Lewis, M., Zettlemoyer, L., Stoyanov, V., 2019. Roberta: A robustly optimized bert pretraining approach. ArXiv abs/1907.11692.

Ma, C., Zhang, W.E., Guo, M., Wang, H., Sheng, Q.Z., 2022. Multi-document summarization via deep learning techniques: A survey. ACM Comput. Surv. URL: https://doi.org/10.1145/3529754, doi:10.1145/3529754. just Accepted.

McBurney, P.W., Liu, C., McMillan, C., 2016. Automated feature discovery via sentence selection and source code summarization. J. Softw. Evol. Process 28, 120–145. URL: https://doi.org/10.1002/smr.1768, doi:10.1002/smr.1768.

McGill, R., Tukey, J.W., Larsen, W.A., 1978. Variations of box plots. The American Statistician 32, 12–16. URL: http://www.jstor.org/stable/2683468.

Mihalcea, R., Tarau, P., 2004. TextRank: Bringing order into text, in: Proceedings of the 2004 Conference on Empirical Methods in Natural Language Processing, Association for Computational Linguistics, Barcelona, Spain. pp. 404–411. URL: https://aclanthology.org/W04-3252.

Moratanch, N., Chitrakala, S., 2016. A survey on abstractive text summarization, in: 2016 International Conference on Circuit, Power and Computing Technologies (ICCPCT), pp. 1–7. doi:10.1109/ICCPCT.2016.7530193.

Moratanch, N., Chitrakala, S., 2017. A survey on extractive text summarization, in: 2017 International Conference on Computer, Communication and Signal Processing (ICCCSP), pp. 1–6. doi:10.1109/ICCCSP.2017.7944061.

Naghshzan, A., 2022. Towards code summarization of apis based on unofficial documentation using nlp techniques. arXiv preprint arXiv:2208.06318 .

Naghshzan, A., Guerrouj, L., Baysal, O., 2021. Leveraging unsupervised learning to summarize apis discussed in stack overflow, in: 2021 IEEE 21st International Working Conference on Source Code Analysis and Manipulation (SCAM), pp. 142–152. doi:10.1109/SCAM52516.2021.00026.

Naghshzan, A., Ratte, S., 2023. Enhancing api documentation through bertopic modeling and summarization. arXiv:2308.09070.

Page, L., Brin, S., Motwani, R., Winograd, T., 1998. The pagerank citation ranking: Bringing order to the web, in: Proceedings of the 7th International World Wide Web Conference, Brisbane, Australia. pp. 161–172. URL: citeseer.nj.nec.com/page98pagerank.html.

Pang, R.Y., Lelkes, A., Tran, V., Yu, C., 2021. AgreeSum: Agreement-oriented multi-document summarization, in: Findings of the Association for Computational Linguistics: ACL-IJCNLP 2021, Association for Computational Linguistics, Online. pp. 3377–3391. URL: https://aclanthology.org/2021.findings-acl.299, doi:10.18653/v1/2021.findings-acl.299.

Papineni, K., Roukos, S., Ward, T., Zhu, W.J., 2002. Bleu: A method for automatic evaluation of machine translation, in: , Association for Computational Linguistics, USA. p. 311–318. URL: https://doi.org/10.3115/1073083.1073135, doi:10.3115/1073083.1073135.

Parnin, C., Treude, C., 2011. Measuring api documentation on the web, in: Proceedings of the 2nd international workshop on Web 2.0 for software engineering, pp. 25–30.

Ponzanelli, L., Bavota, G., Di Penta, M., Oliveto, R., Lanza, M., 2015. Turning the ide into a self-confident programming assistant. .

Qin, Y., Specia, L., 2015. Truly exploring multiple references for machine translation evaluation, in: Proceedings of the 18th Annual Conference of the European Association for Machine Translation, European Association for Machine Translation, Antalya, Turkey. p. . URL: https://aclanthology.org/2015.eamt-1.16.

van Rijsbergen, C., 1979. Information Retrieval.

Shapiro, S.S., Wilk, M.B., 1965. An analysis of variance test for normality (complete samples). Biometrika 52, 591–611. URL: http://www.jstor.org/stable/2333709.

Song, S., Huang, H., Ruan, T., 2019. Abstractive text summarization using lstm-cnn based deep learning. Multimedia Tools Appl. 78, 857–875. URL: https://doi.org/10.1007/s11042-018-5749-3, doi:10.1007/s11042-018-5749-3.

Sridhara, G., Hill, E., Muppaneni, D., Pollock, L., Vijay-Shanker, K., 2010. Towards automatically generating summary comments for java methods, in: , Association for Computing Machinery, New York, NY, USA. p. 43–52. URL: https://doi.org/10.1145/1858996.1859006, doi:10.1145/1858996.1859006.

Student, 1908. The probable error of a mean. Biometrika , 1–25.

Su, D., Xu, Y., Yu, T., Siddique, F.B., Barezi, E., Fung, P., 2020. CAiRE-COVID: A question answering and query-focused multi-document summarization system for COVID-19 scholarly information management, in: Proceedings of the 1st Workshop on NLP for COVID-19 (Part 2) at EMNLP 2020, Association for Computational Linguistics, Online. p. . URL: https://aclanthology.org/2020.nlpcovid19-2.14, doi:10.18653/v1/2020.nlpcovid19-2.14.

Tang, R., Nogueira, R., Zhang, E., Gupta, N., Cam, P., Cho, K., Lin, J., 2020. Rapidly bootstrapping a question answering dataset for covid-19. URL: https://arxiv.org/abs/2004.11339, doi:10.48550/ARXIV.2004.11339.

Tran, N., Tran, H., Nguyen, S., Nguyen, H., Nguyen, T., 2019. Does bleu score work for code migration?, in: 2019 IEEE/ACM 27th International Conference on Program Comprehension (ICPC), pp. 165–176. doi:10.1109/ICPC.2019.00034.

Uddin, G., Khomh, F., 2017. Automatic summarization of api reviews, in: 2017 32nd IEEE/ACM International Conference on Automated Software Engineering (ASE), pp. 159–170. doi:10.1109/ASE.2017.8115629.

Vaswani, A., Shazeer, N., Parmar, N., Uszkoreit, J., Jones, L., Gomez, A.N., Kaiser, L., Polosukhin, I., 2017. Attention is all you need, in: , p. . URL: https://arxiv.org/pdf/1706.03762.pdf.

Wolf, T., Debut, L., Sanh, V., Chaumond, J., Delangue, C., Moi, A., Cistac, P., Rault, T., Louf, R., Funtowicz, M., Davison, J., Shleifer, S., von Platen, P., Ma, C., Jernite, Y., Plu, J., Xu, C., Le Scao, T., Gugger, S., Drame, M., Lhoest, Q., Rush, A., 2020. Transformers: State-of-the-art natural language processing, in: Proceedings of the 2020 Conference on Empirical Methods in Natural Language Processing: System Demonstrations, Association for Computational Linguistics, Online. pp. 38–45. URL: https://aclanthology.org/2020.emnlp-demos.6, doi:10.18653/v1/2020.emnlp-demos.6.

Wu, Y., Hu, B., 2018. Learning to extract coherent summary via deep reinforcement learning, in: Proceedings of the Thirty-Second AAAI Conference on Artificial Intelligence and Thirtieth Innovative Applications of Artificial Intelligence Conference and Eighth AAAI Symposium on Educational Advances in Artificial Intelligence, AAAI Press. p. .

Yang, A., Liu, K., Liu, J., Lyu, Y., Li, S., 2018. Adaptations of ROUGE and BLEU to better evaluate machine reading comprehension task, in: Proceedings of the Workshop on Machine Reading for Question Answering, Association for Computational Linguistics, Melbourne, Australia. pp. 98–104. URL: https://aclanthology.org/W18-2611, doi:10.18653/v1/W18-2611.

Zhang, H., Xu, J., Wang, J., 2019. Pretraining-based natural language generation for text summarization. arXiv preprint arXiv:1902.09243 .